\documentclass[preprint]{aastex}
\usepackage{emulateapj5}
\usepackage{apjfonts}
\usepackage{natbib}

\input psfig.tex

\def\etal{\emph{et al.}\ }

\def\kms{km s$^{-1}$}
\def\apj{ApJ}
\def\aj{AJ}
\def\mnras{MNRAS}
\def\apjs{ApJS}

\def\aa{{A\&A}}

\def\aj{{AJ}}
\def\al{$\alpha$}
\def\bet{$\beta$}
\def\amin{$^\prime$}

\def\apj{{ApJ}}
\def\apjs{{ApJS}}
\def\asec{$^{\prime\prime}$}

\def\etal{{et al. }}

\def\flux{erg s$^{-1}$ cm$^{-2}$}

\def\lamb{$\lambda$}
\def\lum{erg s$^{-1}$}

\def\mnras{{MNRAS}}

\def\pasp{{PASP}}

\def\percm2{cm$^{-2}$}

\def\lax{{$\mathrel{\hbox{\rlap{\hbox{\lower4pt\hbox{$\sim$}}}\hbox{$<$}}}$}}
\def\gax{{$\mathrel{\hbox{\rlap{\hbox{\lower4pt\hbox{$\sim$}}}\hbox{$>$}}}$}}
\def\oii{[\ion{O}{2}]}

\def\oiii{[\ion{O}{3}]}

\slugcomment{To appear in The Astrophysical Journal, September 2002} 
\lefthead{RAVINDRANATH \& HO}
\righthead{REDSHIFTS OF LENSED ARCS IN RX~J1347.5$-$1145}

\begin{document}

\title{Magellan Spectroscopy of the Galaxy Cluster RX~J1347.5$-$1145: Redshift 
Estimates for the Gravitationally Lensed Arcs}

\author{
Swara Ravindranath\altaffilmark{1,2} and
Luis C. Ho\altaffilmark{1} \\
}

\altaffiltext{1}{The Observatories of the Carnegie Institution of Washington, 
813 Santa Barbara St., Pasadena, CA 91101-1292.}

\altaffiltext{2}{Department of Astronomy, University of California, 
Berkeley, CA 94720-3411.}

\setcounter{footnote}{3}

\begin{abstract}
We present imaging and spectroscopic observations of the gravitationally 
lensed arcs in the field of RX~J1347.5$-$1145, the most X-ray luminous
galaxy cluster known. Based on the detection of the \oii\ $\lambda$3727 
emission line, we confirm that the redshift of one of the arcs is $z = 0.806$.  
Its color and \oii\ line strength are consistent with those of distant, 
actively star forming galaxies.  In a second arc, we tentatively identify 
a pair of absorption lines superposed on a red continuum; the lines are 
consistent with \ion{Ca}{2} $\lambda$3933 (K) and \ion{Ca}{2} $\lambda$3968 
(H) at $z = 0.785$.  We detected a faint blue continuum in two additional 
arcs, but no spectral line features could be measured.  We establish lower 
limits to their redshifts based on the absence of \oii\ emission, which we 
argue should be present and detectable in these objects.  Redshifts are also 
given for a number of galaxies in the field of the cluster.
\end{abstract}


\section{Introduction}                   

Gravitational lensing by galaxy clusters serves as a powerful probe of 
cosmological structure.  The lensing phenomenon provides information on both 
the mass distribution of the lensing cluster and the nature of the background 
population of faint field galaxies (e.g., Smail et al. 1993; Fort \& Mellier 
1994). Measurements of cluster mass also place useful constraints on the 
nature of dark matter and the cosmological parameter $\Omega$ (e.g., Mellier, 
Fort, \& Kneib 1993; Fort \& Mellier 1994; Crone, Evrard, \& Richstone 
1994, 1996; White \& Fabian 1995). The arcs that are often seen in deep images 
of galaxy clusters are the sheared and magnified images of faint, background 
galaxies (Paczy\'nski 1987). The amplification provided by gravitational 
lensing enables spectroscopic studies of these faint galaxies, which would 
otherwise be extremely difficult, if not impossible. The spectra of the arcs 
yield information on the redshifts and stellar populations of the lensed 
galaxies (e.g., Smail et al. 1993; B\'{e}zecourt \& Soucail 1997; Ebbels 
et al. 1998; Hall et al. 2000; Campusano et al. 2001). The redshifts of the 
arcs constrain models of the the cluster potential and provide robust 
estimates of the total cluster mass.

RX~J1347.5$-$1145, at a redshift of 0.451, is the most luminous X-ray cluster 
known, with an X-ray luminosity in excess of 10$^{45}$ \lum\ (Schindler et al. 
1995, 1997; Ettori, Allen, \& Fabian 2001). The mass estimates based on its 
X-ray properties (Schindler et al. 1995), the Sunyaev-Zel'dovich effect 
(Pointecouteau et al. 1999), weak-lensing models (Fischer \& Tyson 1997), and 
strong-lensing models (Sahu et al. 1998; Cohen \& Kneib 2002) have yielded 
discrepant results for the total mass of this cluster. Cohen \& Kneib (2002) 
speculate that we may be witnessing the merging of two clusters along a 
direction perpendicular to our line of sight. 

Strong constraints can be placed on the cluster mass based on the redshifts of 
lensed background galaxies. Schindler et al. (1995) discovered an arc system in 
RX~J1347.5$-$1145; it comprises of two bright arcs located $\sim$35\asec\ from 
the central dominant galaxies, at diametrically opposite points along the 
North-South direction.  This was confirmed by the observations of Fischer \& 
Tyson (1997).  For the bright northern arc (Arc~1), Sahu et al. (1998) 
reported the detection of an emission line plausibly identified with 
\oii\ $\lambda$3727 at a redshift of 0.81.  The bright southern arc (Arc~4) 
showed a faint blue continuum but no spectral line features. 
The high-resolution {\it Hubble Space Telescope}\ images of Sahu et al. (1998) 
also revealed three additional arcs in this cluster (Arcs 2, 3, and 5). 
Figure~1 shows an $R$-band image of the central region of the cluster
with identification for the five arcs.  

This paper presents new photometry and spectroscopy of the arc system 
in RX~J1347.5$-$1145. We confirm the emission-line redshift 
previously reported for Arc~1, report the tentative detection of an 
absorption-line redshift for Arc~2, and give limits on the redshifts of 
Arcs 3 and 4.  These measurements are in good agreement with the
predictions from the lensing models of this cluster.  We also give redshifts 
for a number of galaxies in the field of the cluster.

\section{Observations and Data Reductions}        

In preparation for the spectroscopic observations, on 2001 February 20 UT we 
acquired relatively deep $V$ (2$\times$1200~s) and $I$ (2$\times$900~s) images 
of RX~J1347.5$-$1145 using the 2.5-m du~Pont telescope at Las Campanas 
Observatory. The CCD was a thinned Tektronics $2048\times 2048$ chip 
with a pixel scale of 0\farcs26, a gain of 3.0 $e^{-}$ per ADU, and a read 
noise of 7 $e^{-}$.  The data were taken under photometric, sub-arcsecond 
($\sim$0\farcs8) conditions.  The images were used for photometry of the arcs 
and to obtain accurate relative astrometry for the preparation of the slit 
mask for the spectroscopic observations.  Photometric and astrometric 
calibrations were achieved by observing the open cluster M~67 (Montgomery, 
Marschall, \& Janes 1993).

We obtained spectra of the brightest four arcs (Arcs 1--4 in the notation of 
Sahu et al. 1998) on 2001 May 15--16 UT, using the multislit LDSS-2 
spectrograph 
(Allington-Smith et al. 1994) mounted on the 6.5-m Magellan~I (Baade) telescope 
at Las Campanas.  We did not include Arc~5 because it partly overlaps with Arc~1
along the dispersion direction.  The 
%
\vskip 0.3cm
\psfig{file=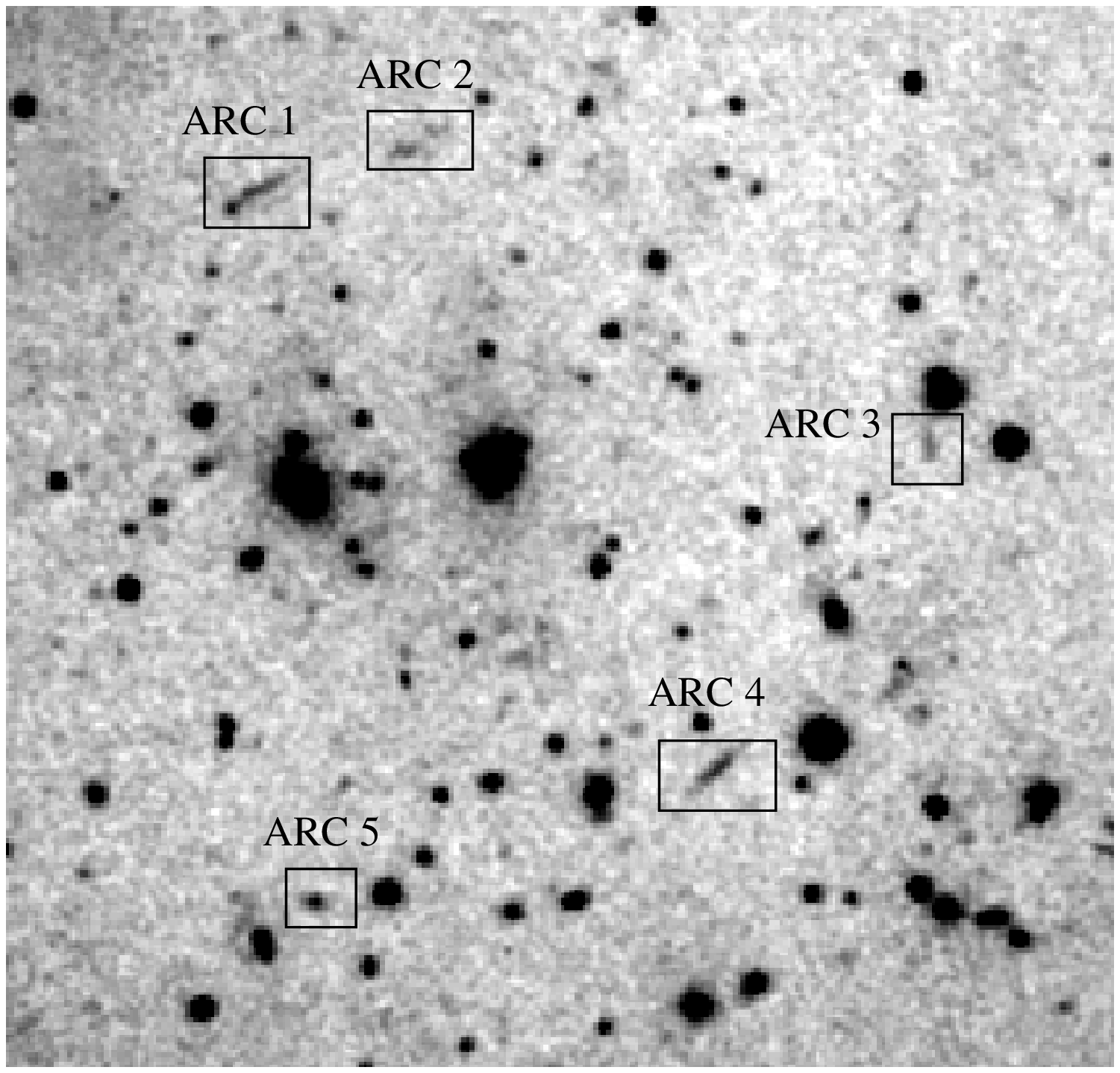,width=8.5cm}
\figcaption[fig1.ps]{The $R$-band acquisition image of RX~J1347.5$-$1145
obtained using LDSS-2 on the Baade telescope. The gravitationally lensed
arcs are identified using the notation of Sahu et al. (1998). The field shown
here covers a 1.8\amin $\times$ 1.8\amin\ region centered around the dominant
cluster galaxies. North is to the top and East is to the left.
\label{fig1}}
\vskip 0.3cm
%
\noindent
300 line~mm$^{-1}$ grism blazed at 6000 
\AA\ covered the spectral range $\sim$4000 to 8000 \AA\ with a dispersion 
of 5.3 \AA\ pixel$^{-1}$.  The two-dimensional spectra were recorded on a SITe 
$2048\times 2048$ CCD, which has a spatial resolution of 0\farcs38 
pixel$^{-1}$.  In order 
to maximize the light input from the faint, low surface brightness  
arcs ($\mu_V \approx 25-26$ mag arcsec$^{-2}$), we chose the width
of the slits to be 2\asec.  This gave a full width at half maximum (FWHM)
spectral 
resolution of 25~\AA.  For simplicity, all the slits were oriented along the 
East-West direction, at the expense that for most of the arcs only a portion of 
each object intersected with the slit.  The slits were required to have a 
minimum length of 10\asec\ in order to have enough area for sky subtraction. 
In addition to the arcs, the mask also contained slits for a number of 
galaxies that were likely to be associated with the cluster.  
The total integration time was 21,400~s 
($\sim$6~hr), split into seven roughly equal exposures.  The sky conditions 
were clear, the atmospheric seeing varied between 0\farcs6 and 0\farcs9, and 
the airmass ranged from 1.0 to 1.65.  

The basic data reductions were carried out using the IRAF\footnote{IRAF is 
distributed by the National Optical Astronomy Observatories, which are operated 
by the Association of Universities for Research in Astronomy, Inc., under 
cooperative agreement with the National Science Foundation.} package. The 
du~Pont images were bias-subtracted, flat-fielded with twilight 
skyflats, aligned, and then median combined after cosmic-ray rejection.  
Photometry of the arcs and galaxies was performed using the SExtractor 
software (Bertin \& Arnouts 1996), 
which uses elliptical apertures to compute magnitudes within an 
isophotal radius that is twice the first-moment radius (Kron 1980). The maximum 
radius used to compute the first-moment radius corresponds to an isophotal 
value that is 1.5 $\sigma$ of the background. 
The formal errors on the magnitudes derived from SExtractor are $\sim$
0.3 mag.  The magnitudes were corrected for 
atmospheric extinction and transformed to the standard Johnson system using
calibrations derived from photometry of the standard stars in M~67. 

The spectroscopic frames were corrected for the overscan, flat-fielded with 
domeflats, aligned, and then median combined after cosmic-ray rejection.  
Accurate sky subtraction proved to be challenging especially for the 
lensed arcs.  This is in part due to the extremely low surface brightness 
of the arcs and to the fact that the 
spectra are slightly tilted.  After some experimentation, we found that 
two-dimensional background subtraction gave the most robust results.  The 
background was estimated by fitting a third-order Chebyshev polynomial along 
the columns using all the rows excluding those that contain the object 
spectrum.  We extracted one-dimensional spectra by summing the central 
10 rows ($\sim$4\asec); this extraction width was determined empirically to 
give the best signal-to-noise ratio (S/N).  We established the wavelength scale 
by fitting a third-order polynomial to unblended emission lines of He and Ne
in the comparison lamp spectra. Finally, the relative fluxes of the spectra 
were calibrated using longslit observations of the standard stars LTT~7987 and
LTT~9491 (Stone \& Baldwin 1983).  Note that we did not attempt to perform
absolute spectrophotometry because our primary goal was to obtain redshifts
for the arcs and galaxies. 

\section{Properties of the Arcs}           

Table~1 summarizes the integrated magnitudes and colors obtained from the 
broad-band images. Figure~2 shows the final spectra for the lensed arcs in
RX J1347.5$-$1145.  For each arc, we present 
both the two-dimensional background-subtracted spectrum (in greyscale) and the 
extracted, calibrated one-dimensional spectrum.   

The spectrum of Arc~1 shows an unresolved emission line centered at $\sim$6728 
\AA.  The emission-line knot is clearly visible in the two-dimensional
sky-subtracted image, and it is unmistakable in each of the sub-exposures. This 
line was already reported by Sahu et al. (1998), who argued that it is likely 
to be redshifted \oii\ $\lambda$3727.  Adopting this interpretation gives a 
redshift of 0.806. The line has an equivalent width (EW) of 85$\pm$10 \AA, 
where the error bars reflect uncertainties in the placement of the continuum 
level and in whether the line flux is measured by direct integration or profile 
fitting. We note that although our absolute fluxes are not reliable because of 
slit losses, the EW measurements are likely to be more robust, so long as 
there are not large spatial variations of the line-emitting regions within the 
galaxy. From the photometry, we measure a moderately red color of 
$V-I \approx 1.6$ mag, similar to the color $B_{J}-R$ = 1.1 mag given by 
Fischer \& Tyson (1997).

Arc~2, for which there have been no previous spectroscopic observations, 
exhibits a pair of absorption lines at 7017 and 7084 \AA\ superposed 
on a relatively red continuum.  The color derived from the images is also 
rather red, with $V-I \approx 2.7$ mag.  The lines fall on a fairly clean 
part of the spectrum and do not appear to be adversely affected by sky 
subtraction.  

We clearly detected the continuum in Arcs~3 and 4, but the spectra do not 
reveal any distinct emission or absorption features that can provide a direct 
redshift measurement. The feature near 5200 \AA\ in the spectrum of Arc 4 
appears not to be real.  It is not seen in the background-subtracted 
two-dimensional image and may have resulted from the addition of significant
residuals within the extraction aperture .  The spectrum of Arc~3 is 
substantially noisier than that of Arc~4, for two reasons.  First, Arc~3 is 
$\sim$1.5 mag fainter.  And second, the slit was oriented in an especially 
unfavorable position angle with respect to the object; the length of the arc 
runs North-South (see Fig.~1), exactly orthogonal to the slit.  Arcs~3 and 4 
are significantly bluer than Arcs~1 and 2.  
\begin{figure*}[t]
\psfig{file=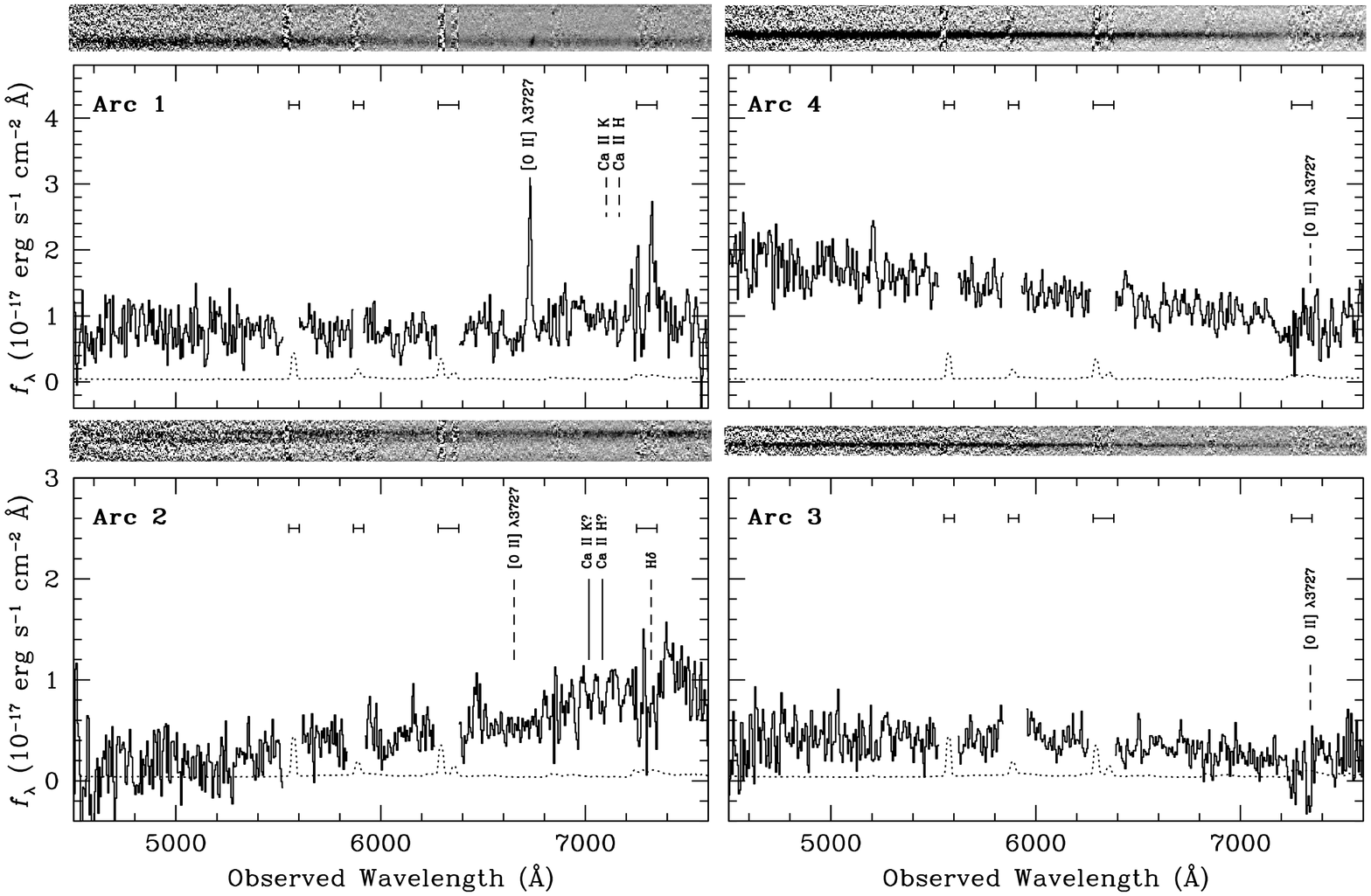,width=18.8cm,angle=90}
\figcaption[fig2.ps]{Spectra of the gravitational arcs in RX~J1347.5$-$1145.
For each arc, its two-dimensional sky-subtracted spectrum (in greyscale)
is shown along with the one-dimensional extracted spectrum; both cover the same
wavelength range (4500$-$7600\AA).  The dotted line in each panel shows
the sky spectrum, scaled by a factor of 0.01. The gaps in the object spectrum
corresponds the positions of strong sky lines (indicated by bars) where
imperfect sky subtraction leaves large residuals. The spectral features used
to infer the redshifts are marked with solid lines, while the dashed lines
mark the positions of other features expected for the inferred redshifts.
In the spectra of Arc~3 and 4, we marked the position of \oii\ $\lambda$3727
expected for $z = 0.97$, as predicted by Allen et al. (2002).
\label{fig2}}
\end{figure*}
\noindent
We measure $V-I \approx 0.41$ and 
0.58 mag for Arcs~3 and 4, respectively; for comparison, Fischer \& Tyson 
(1997) give $B_{J}-R$ = 0.4 mag for Arc~4.


\section{Redshifts for Galaxies in the Field of RX J1347.5$-$1145}   

We obtained spectra for 22 galaxies in the field of the cluster; Table 2 gives 
redshifts and basic photometric data for 21 of these.  The galaxy redshifts 
were determined using the cross-correlation technique of Tonry \& Davis (1979) 
as implemented in the {\it xcsao}\ task in IRAF.  Cross-correlation templates 
were taken from the spectrophotometric atlas of galaxies of Kennicutt (1992a). 
These templates cover the wavelength range 3650$-$7100 \AA\ at a resolution of 
5$-$8 \AA.  We correlated each galaxy spectrum with a set of seven 
template spectra chosen to represent a range of spectral types. The 
templates include NGC 3379 (E1), NGC 4472 (E2), NGC 4889 (E4), NGC 3921 
(S0/a), NGC 3627 (Sb), NGC 6764 (Sbc), and NGC 6240 (I0). NGC 3921 has a 
post-starburst spectrum showing a mixed population with strong Balmer 
absorption lines typical of A stars along with K-giant features. Both 
absorption-line and emission-line templates were used for redshift 
determination, depending on the spectrum of the observed galaxy. 

%
\begin{figure*}[t]
\centerline{\psfig{file=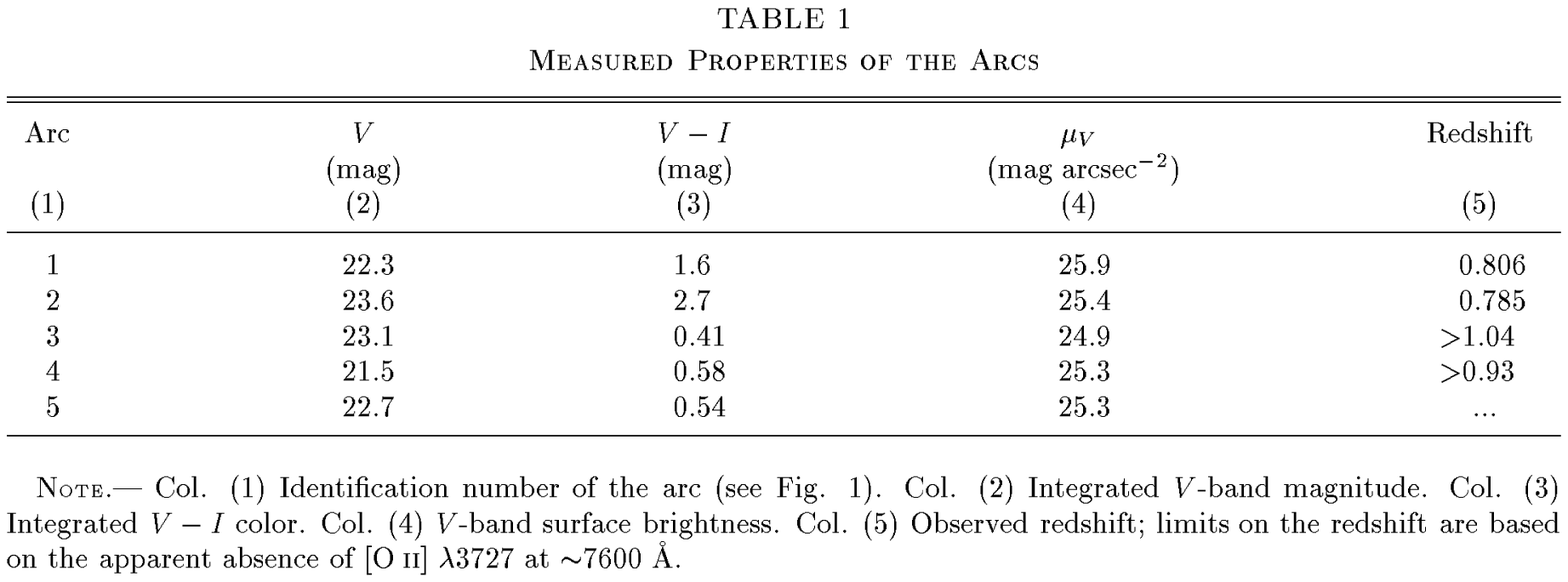,width=13.5cm,angle=0}}
\end{figure*}
%


The two main contributions to the errors in the measured redshifts arise
from template mismatch and errors in the wavelength calibration. The errors 
due to template mismatch can be estimated from the dispersion in the final 
redshifts resulting from different input templates.  For redshifts based on 
absorption-line templates, these errors are typically $\sim$350 km s$^{-1}$. 
The errors are significantly smaller when the redshifts are determined using 
emission-line templates, with $\sim$100 km s$^{-1}$ being a typical value. 
The rms error in the wavelength calibration of our observed galaxy spectra was 
found to be $\sim$5 \AA. Adding the two contributions in quadrature, we 
estimate a total error of $\sim$ 0.0014 for the derived redshifts.  In 
addition, there can be errors in the velocity zeropoints due 
to wavelength calibration errors of the template galaxy spectra. However, these 
are of the order of few tens of km s$^{-1}$ (Yee, Ellingson, \& Carlberg 1996),
and we have not included them in our error estimation.   

Our redshift estimates show excellent agreement with the results 
reported by Cohen \& Kneib (2002) in the case of four galaxies
common to both studies. We assign cluster membership based on 
the velocity distribution presented by Cohen \& Kneib (2002) 
for 47 spectroscopically confirmed cluster members.
Five galaxies in our study have redshifts close to the redshifts for the 
two central cD galaxies given by Cohen \& Kneib (2002), and thus are likely to 
be cluster members. Among these, the galaxy with $z$ = 0.431 is at the tail 
end of the velocity distribution, but we include it as a cluster member 
because its red $V-I$ color is similar to that of the other cluster galaxies.  
Another probable cluster member is a galaxy at redshift $z$ = 0.426
that lies close to the range spanned by the velocity distribution. This galaxy 
shows an absorption-line spectrum, but its $V-I$ color is relatively blue 
compared to the other cluster members.  Only one of the galaxies among the 
likely cluster members, G47195$-$4344, shows an emission-line spectrum; it 
is located at the outskirts of the cluster, $\sim 3^{\prime}$ from the central 
cD galaxies. A similar cluster member candidate was reported by 
Cohen \& Kneib (2002); C47229\_4519 has a relatively blue color and is the 
only cluster galaxy in their sample other than the central cD galaxy (which 
hosts an active nucleus) to show the \oii\ $\lambda$3727 emission line.  

%
\begin{figure*}[t]
\centerline{\psfig{file=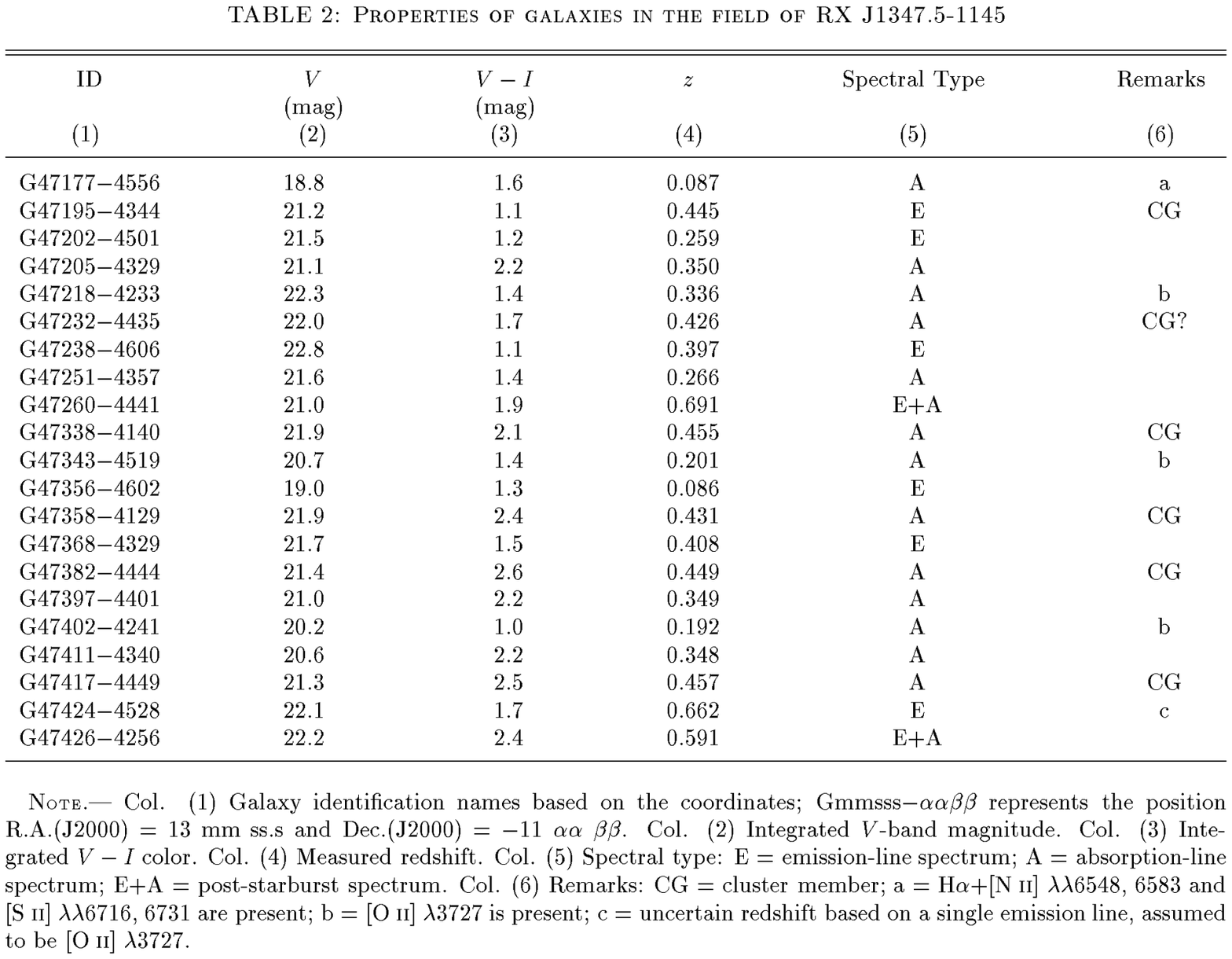,width=13.5cm,angle=0}}
\end{figure*}
%

\section{Discussion and Summary}          

Our primary aim is to determine or place limits on the redshifts of the
arcs in RX~J1347.5$-$1145.  The redshift for Arc~1 is relatively secure. 
Consistent with the study of Sahu et al. (1998), we detected a strong 
emission line at $\sim$6728 \AA\ whose most likely identification is 
\oii\ $\lambda$3727 at $z = 0.806$.  As discussed by Sahu et al. (1998), 
Ly$\alpha$ $\lambda$1216 or \ion{C}{4} $\lambda$1550 can be ruled out based 
on the photometric colors: the Lyman limit would render the $B_{J}-R$ 
color much redder than that reported by Fischer \& Tyson (1997).  More 
directly, our spectrum, which extends to $\sim$4500 \AA, shows no sign of 
any continuum decrement.  Strong optical lines such as H\bet, 
\oiii\ \lamb\lamb4959, 5007, or H\al\ can be ruled out trivially by the 
redshift (0.451) of the lensing cluster.

The detection of the \oii\ line allows us to deduce a few basic properties 
concerning Arc~1. The measured \oii\ EW of $\sim$75--95 \AA\ compares well 
with values previously found in gravitationally lensed arcs seen toward 
other galaxy clusters (Ebbels et al. 1998).  It is somewhat larger than in 
typical nearby late-type galaxies (EW $\approx$ 50 \AA; Kennicutt 1992b), but 
lies in the upper end of the EW distribution seen in distant, faint galaxy 
samples (e.g., Colless et al. 1990; Hammer et al. 1997).  The observed $V-I$ 
color is also consistent with that expected for the faint blue galaxy 
population at $z \approx 0.8$ (Forbes et al. 1996).  Although slit losses 
prevent us from measuring accurate total fluxes, we can use the observed 
\oii\ emission-line flux to set a lower limit to the formation rate of massive 
(ionizing) stars.  Kennicutt (1998) gives the following empirical relation 
between \oii\ luminosity ($L_{\rm [O~II]}$) and star formation rate (SFR):
\begin{equation}
{\rm SFR} (M_{\odot} \, {\rm yr^{-1}}) = (1.4 \pm 0.4) \times 10^{-41} \,
L_{\rm [O~II]} \,\, ({\rm erg \, s^{-1}}).
\end{equation}
\vskip 0.3cm
\noindent
This derivation assumes a Salpeter (1955) stellar initial mass function and 
solar metallicity.  For an observed $F_{\rm [O~II]} > 5.5\times10^{-16}$ \flux\ 
and cosmological parameters $H_0$ = 75 \kms\ Mpc$^{-1}$, 
$\Omega_{\rm m} = 0.3$, and $\Omega_{\lambda} = 0.7$, we obtain SFR $>$ 24 
$M_{\odot}$~yr$^{-1}$.  The above calculation accounts for a Galactic 
extinction of $A_B = 0.268$ mag (Schlegel, Finkbeiner, \& Davis 1998) corrected 
using the extinction law of Cardelli, Clayton, \& Mathis (1989).  To obtain the 
{\it intrinsic}\ SFR, we need to know the flux magnification factor due to the 
lensing.  According to the lensing model of Allen, Schmidt, \&  Fabian (2002), 
the magnification factor of Arc~1 is 7.7 (R.~W. Schmidt, private communication).
Thus, the true lower limit to the SFR for Arc~1 is $\sim$3 
$M_{\odot}$~yr$^{-1}$.  This is comparable to the level of star formation 
activity in nearby gas-rich spiral galaxies (e.g., Kennicutt 1983) but is 
significantly lower than those obtained for the arcs in Abell~2218 (Ebbels et 
al. 1996) and Abell~2390 (B\'{e}zecourt \& Soucail 1997; L\'emonon et al. 
1998).  It is unclear whether slit losses alone can make up for the difference.

For Arc~2, we tentatively identify the pair of absorption lines at 7017 
and 7084 \AA\ with \ion{Ca}{2} $\lambda$3933 (K) and \ion{Ca}{2} $\lambda$3968 
(H) at a redshift of 0.785.  This interpretation is plausible considering (1) 
the redness of the continuum, which is suggestive of an old stellar population, 
and (2) the absence of any strong emission lines blueward of the absorption 
lines.  An old stellar population should show a more prominent 4000 \AA\ break 
than observed, but the shape of the spectrum redward of $\sim$7200 \AA\ is too 
uncertain (due to telluric molecular absorption bands and residuals from 
subtraction of sky lines) to be definitive on this point.  For an elliptical 
galaxy at $z \approx 0.8$, the observed $V-I$ color of 2.7 mag corresponds to a 
present-day restframe $V-I \approx 1.2$ mag (Poggianti 1997), consistent 
with the colors of local elliptical galaxies (e.g., Fukugita, Shimasaku, \& 
Ichikawa 1995).  Allen et al. (2002) recently used {\it Chandra}\ 
data to refine the mass model for RX~J1347.5$-$1145.  Adjusting their model to 
reproduce the redshift measurement of Arc~1 by Sahu et al. (1998), Allen et al. 
(2002) predict the redshifts of the other arcs.  For Arc~2, they give 
$z = 0.75 \pm 0.05$, in excellent agreement with our value.

Allen et al. (2002) predict $z = 0.97 \pm 0.05$ for Arcs~3 and 4.  Since 
both of these objects have featureless continua very similar in shape to 
that of Arc~1 --- indeed, they are {\it bluer}, suggesting an even younger 
stellar population --- it is reasonable to expect that nebular emission 
should be present at a comparable, if not even greater, strength.  However, no 
significant emission feature is discernible at $\sim$7340 \AA, the expected 
location of \oii\ at $z = 0.97$.  Unfortunately, the quality of the spectrum in 
the region $\sim$7300--7400 \AA\ is degraded by imperfect removal of the 
OH sky lines (see, e.g., Osterbrock \& Martel 1992).  Nevertheless, a narrow 
emission feature with a strength comparable to that of the \oii\ line in 
Arc~1, and perhaps even a factor 2--3 weaker, would almost certainly have been 
detected in Arc~4, since the two objects have virtually identical continuum 
levels.  These arguments suggest that the redshift of Arc~4 is greater than 
1.04, where we have taken the upper limit of our bandpass to be 7600 \AA.  This 
redshift limit does not appear to be in serious conflict with the predicted 
value.

The faintness of Arc~3 makes its spectrum highly uncertain at the red end, and 
we limit our discussion to wavelengths \lax 7200 \AA, where the continuum is 
clearly detected and not severely affected by systematic effects.  The absence 
of any emission features with equivalent widths greater than $\sim$10 \AA\ 
suggests that the redshift is likely to be larger than 0.93.  This is 
consistent with the value predicted by Allen et al. (2002).

\acknowledgements
This work is funded by NASA LTSA grant NAG 5-3556 and by NASA grants 
HST-AR-07527.03-A and HST-AR-08361.02-A from the Space Telescope Science 
Institute (operated by AURA, Inc., under NASA contract NAS5-26555). We are 
grateful to Paul Martini for obtaining the broad-band images of the cluster. 
We thank Andrew McWilliam, Patrick McCarthy, and Daniel Kelson for helpful 
discussions on the spectroscopic reductions.  Robert Schmidt kindly 
communicated the magnification factors for the arcs.  An anonymous referee 
gave helpful suggestions for improving the paper.  


\end{document}